\documentclass[a4paper]{revtex4}
\usepackage{graphicx}
\usepackage{fancyhdr}
\usepackage{amsmath}
\usepackage{color}

\begin{document}

\title{
Higgs boson couplings as a probe of new physics }

%

\author{Kei Yagyu}
\affiliation{School of Physics and Astronomy, University of Southampton, Southampton, SO17 1BJ, United Kingdom}

\begin{abstract}
 
Precise measurements of various coupling constants of the 125 GeV Higgs boson $h$ are one of the most important and solid methods 
to determine the structure of the Higgs sector. 
If we find deviations in the $h$ coupling constants from the standard model predictions, 
it can be an indirect evidence of the existence of additional Higgs bosons in non-minimal Higgs sectors. 
Furthermore, we can distinguish non-minimal Higgs sectors by measuring a 
pattern of deviations in various $h$ couplings. 
In this talk, we show patterns of the deviations in several simple non-minimal Higgs sectors, 
especially for the gauge $hVV$ and Yukawa $hf\bar{f}$ couplings. 
This talk is based on the paper~\cite{fingerprint}. 

\end{abstract}

\maketitle
\thispagestyle{fancy}


\section{Introduction}

The experimental results provided by CERN Large Hadron Collider (LHC)~\cite{LHC2,LHC3,LHC4} with the 7 and 8 TeV energies 
strongly suggested us that there is at least one isospin doublet Higgs field. 
Although this situation can be most economically explained in  
the minimal Higgs sector with only one doublet assumed in the standard model (SM), 
there still remain a number of questions regarding the Higgs physics. 
First, we do not know what is the true structure of the Higgs sector, i.e.,  
the number of scalar multiplets, those quantum numbers, and symmetries in the potential. 
Second, what is the relationship between the structure of the Higgs sector and the phenomena which cannot be explained in the SM, e.g., 
neutrino masses, the existence of dark matter and baryon asymmetry of the Universe?
Third, it is interesting to consider the decoupling property of the Higgs sector, i.e., whether the Higgs sector has a non-decoupling or decoupling property. 
Finally, what is the mass scale of the second Higgs boson?
In addition to the above mentioned questions, 
there are many unknown things about the Higgs sector. 
Therefore, finding the answer for these questions, we obtain deeper understanding of elementary particle physics. 

On the other hand, so far, many new physics models beyond the SM have been considered in various physics motivations. 
In most of the cases, the Higgs sector is extended from the minimal form. 
For example, isospin singlet scalar fields are introduced in models with gauged $U(1)_{B-L}$ symmetry~\cite{BL}. 
In supersymmetric SMs, we need at least two doublet Higgs fields to give masses of SM fermions. 
The multi-doublet structure is also motivated to obtain an additional source of CP-violation~\cite{CPV} in the Higgs sector which
is necessary to realize a successful scenario of the electroweak baryogenesis~\cite{ewbg}. 
Furthermore, an isospin triplet field is introduced, e.g., in the type-II seesaw model~\cite{typeII} and left-right symmetric models~\cite{left-right}. 
Because a property of the Higgs sector strongly depends on new physics models, 
{\it Higgs is a probe of new physics. }

How can we determine the true structure of the Higgs sector?
There are basically two methods, namely, (i) a discovery of additional Higgs bosons, and (ii) measuring deviations in Higgs bosons couplings. 
Although both methods are important to reconstruct the shape of the Higgs sector, 
in this talk, we focus on the second one. 
In general, when we consider non-minimal Higgs sectors, Higgs boson couplings can be modified due to a mixing among neutral Higgs bosons and VEVs among Higgs 
multiplets. These effects can happen at the tree level. 
In addition, Higgs boson couplings can also deviate from the SM predictions by radiative corrections. 

In this talk, 
we first review basic experimental constraints to restrict the form of Higgs sector. We then discuss deviations in the Higgs boson couplings in 
several simple non-minimal Higgs sectors. 
Finally, we conclude this talk, and briefly introduce several works for radiative corrections to the Higgs boson couplings. 

\section{Basic Constraints}

There are two important experimental requirements to constrain the form of the Higgs sector, which come from 
the electroweak rho parameter and the flavor changing neutral current (FCNC). 

First, we discuss the constraint from the rho parameter. 
In the general Higgs sector with a scalar field with the isospin $T_i$, the hypercharge $Y_i$ (the electric charge $Q$ is given by $Q=T^3+Y$) and the VEV $v_i$, 
the rho parameter is given at the tree level by~\cite{HHG}
\begin{align}
\rho_{\text{tree}} = \frac{\sum_i v_i^2[T_i(T_i+1)-Y_i^2]}{\sum_i 2Y_i^2v_i^2}.  \label{eq1}
\end{align}
From the experimental data $\rho_{\text{exp}}=1.0004^{+0.0003}_{-0.0004}$~\cite{PDG}, 
Higgs sectors with $\rho_{\text{tree}}=1$ seem to be a natural choice.
Combinations of $T_i$ and $Y_i$ which do not change $\rho_{\text{tree}}=1$ is obtained from Eq.~(\ref{eq1}) as
\begin{align}
T_i(T_i+1)=3Y_i^2.  \label{rho}
\end{align}
When $T_i$ and $Y_i$ are given to be half-integers, we obtain $(T_i,Y_i)=(0,0),~(1/2,1/2),~(3,2), \dots$. 
If we allow to align VEVs of several Higgs multiplets, we can keep $\rho_{\text{tree}}=1$. 
The minimal realization of such an alignment is seen in the Georgi-Machacek (GM) model~\cite{GM}, where 
one $Y=1$ and one $Y=0$ triplets are introduced in the Higgs sector. We can take two triplet VEVs so as to 
keep the approximate custodial $SU(2)_V$ symmetry in the kinetic term as it is realized in the SM. 
Recently, in Ref.~\cite{Logan}, the alignment mechanism in the GM model has been generalized in larger Higgs representations. 
If we introduce a Higgs multiplet without satisfying Eq.~(\ref{rho}), and do not impose an alignment of VEVs, 
a magnitude of such a VEV should be tuned to be quite small as compared to the doublet VEV. 
For example, if we consider a Higgs sector with a doublet with $Y=1/2$ and a triplet with $Y=1$, then the triplet VEV is constrained to be smaller than 
about 3.5 GeV at 95\% CL~\cite{yokoya}. 

Another important constraint to restrict the form of the Higgs sector comes from suppression of the FCNCs. 
For example, the $B_0$-$\bar{B}_0$ mixing is successfully suppressed in the SM, because of the Glashow-Iliopoulos-Maiani mechanism, and 
it appears via the one loop W boson exchanging box diagrams. 
However, when we consider multi doublet models, this virtue is lost, because 
in general, neutral Higgs bosons contribute to FCNC processes at the tree level. 
This can be clearly seen in two Higgs doublet models (THDMs) as the simplest example. 
Let us consider, e.g., the down-type Yukawa interaction in THDMs which is given by (flavor indices are neglected)
\begin{align}
{\cal L}_Y = \bar{Q}_L (\tilde{Y}_{1}\Phi_1 + \tilde{Y}_{2}\Phi_2)d_R +\text{h.c.},  \label{yuk}
\end{align}
where $\Phi_1$ and $\Phi_2$ are the Higgs doublets, and $Q_L~(d_R)$ is the left (right) handed quark doublet (singlet). 
In the so-called Higgs basis, the Yukawa interactions are rewritten as 
\begin{align}
{\cal L}_Y = \bar{Q}_L (Y_{1}\Phi + Y_{2}\Psi)d_R +\text{h.c.},  \label{yuk2}
\end{align}
where $\Phi$ has the VEV $v\sim 246$ GeV, three Nambu-Goldstone bosons, and a neutral Higgs boson, while $\Psi$ does not have a VEV but includes 
physical Higgs states which are not necessarily mass eigenstates. 
The original basis and the Higgs basis are related by the orthogonal transformation as 
\begin{align}
\begin{pmatrix}
\Phi_1 \\
\Phi_2
\end{pmatrix}=\begin{pmatrix}
\cos\beta & -\sin\beta \\
\sin\beta & \cos\beta
\end{pmatrix}
\begin{pmatrix}
\Phi \\
\Psi
\end{pmatrix},~~\text{with}~~\tan\beta =\langle \Phi_2^0 \rangle/\langle \Phi_1^0 \rangle. 
\end{align}
The two Yukawa matrices $\tilde{Y}_a$ and $Y_a$ are then related with each other by 
\begin{align}
Y_1 = \tilde{Y}_1\cos\beta  +\tilde{Y}_2 \sin\beta ,\quad
Y_2 = -\tilde{Y}_1 \sin\beta  +\tilde{Y}_2 \cos\beta  .  \label{mat}
\end{align}
The mass matrix for the down-type quarks is obtained from the first term of Eq.~(\ref{yuk2}), so that 
the left and right handed quarks should be rotated so as to diagonalize the matrix $Y_1$. 
However, in general, the matrix $Y_2$ given in Eq.~(\ref{mat}) is not diagonalized at the same time, because there is no reason 
why $Y_1$ and $Y_2$ are proportional to each other. 
Therefore, flavor violating interactions are derived from the second term of Eq.~(\ref{yuk2}). 
Such a situation can be avoided if only one of the doublets couple to the down-type quarks. In other words, 
when either a $\tilde{Y}_1$ or $\tilde{Y}_2$ term is forbidden, then $Y_1$ is proportional to $Y_2$ as seen in Eq.~(\ref{mat}). 
By applying the similar procedure to the up-type quark and lepton sectors, 
we can define four independent types of Yukawa interactions~\cite{Barger,Grossman} so-called Type-I, Type-II, Type-X and Type-Y. 
According to Ref.~\cite{typeX}, the definition of four types are given in Table~\ref{types}. 
The Yukawa interaction terms are then given by 
\begin{align}
{\cal L}_Y = \bar{Q}_L \frac{M_d}{v}(\Phi + \sqrt{2}\xi_d\Psi)d_R 
+\bar{Q}_L \frac{M_u}{v}(i\sigma_2\Phi^* + \sqrt{2}\xi_u(i\sigma_2)\Psi^*)u_R
+\bar{L}_L \frac{M_e}{v}(\Phi + \sqrt{2}\xi_e\Psi)e_R
+\text{h.c.},  \label{yuk3}
\end{align}
where $M_f$ $(f=u,d,e)$ are the fermion mass matrices and $\xi_f$ factors are given in Table~\ref{types} 
The simplest way to realize such a situation is to impose a discrete $Z_2$ symmetry~\cite{GW}, where doublet fields are transformed as 
$\Phi_1\to +\Phi_1$ and $\Phi_2\to  -\Phi_2$, and the types of Yukawa interactions are determined by fixing the $Z_2$ charges of fermions~\footnote{As alternative possibilities, we can impose 
a larger discrete symmetry such as $S_3$~\cite{S3}, or an $U(1)$ symmetry~\cite{Omura}. }.

\begin{table}[ht]
\begin{center}
\caption{Definition of the four types of Yukawa interactions. For example, in the Type-II, $\Phi_1$ ($\Phi_2$)
couples to down-type quarks ($d$) and charged leptons $e$ (up-type quarks $u$). The $\xi_f$ factors in Eq.~(\ref{yuk3}) are also shown in each of types. }
\begin{tabular}{|l|c|c|c|c|}
\hline  & Type-I & Type-II & Type-X & Type-Y \\
\hline $\Phi_1$ & -          & $d,~e$ & $e$ & $d$\\
\hline $\Phi_2$ & $u,~d,~e$  & $u$ & $u,~d$ & $u,~e$  \\
\hline $(\xi_u,\xi_d,\xi_e)$ & $(\cot\beta,\cot\beta,\cot\beta)$  & $(\cot\beta,-\tan\beta,-\tan\beta)$ & $(\cot\beta,\cot\beta,-\tan\beta)$ & $(\cot\beta,-\tan\beta,\cot\beta)$  \\
\hline
\end{tabular}
\label{types}
\end{center}
\end{table}


\section{Modifications of Higgs boson couplings}

In this section, we discuss deviations in the coupling constants of the SM-like Higgs boson ($h$)
with the weak gauge bosons $hVV$ ($V=W,Z$) and fermions $hf\bar{f}$ in several simple extended Higgs sectors. 
In order to express the deviation in the Higgs boson couplings, it is convenient to introduce the quantities so-called scaling factors defined as follows:
\begin{align}
\kappa_V^{} = \frac{g_{hVV}^{}}{g_{hVV}^{\text{SM}}}, \quad  
\kappa_f^{} = \frac{g_{hf\bar{f}}}{g_{hf\bar{f}}^{\text{SM}}},
\end{align}
where $g_{hf\bar{f}}$ and $g_{hVV}$ denote the $hf\bar{f}$ and $hVV$ couplings, respectively, and those with ``SM'' stand for corresponding SM values. 
These scaling factors have been derived from the Higgs boson search data collected by 
ATLAS and CMS Collaborations with $\sqrt{s}=$ 7 and 8 TeV and 25 fb$^{-1}$ of the integrated luminosity~\cite{LHC2,LHC3,LHC4}. 
From the two parameter fit analysis based on Ref.~\cite{Handbook}, 
current data give
\begin{align}
&\kappa_V^{} = 1.15\pm 0.08, \quad \kappa_f^{} = 0.99^{+0.08}_{-0.15},\quad \text{ATLAS~\cite{LHC2}}, \label{hc1}\\
&\kappa_V^{} = 1.01\pm 0.07, \quad \kappa_f^{} = 0.87^{+0.14}_{-0.13},\quad \text{CMS~\cite{LHC4}}, \label{hc2}
\end{align}
where universal scaling factors, i.e., 
$\kappa_F^{} = \kappa_t=\kappa_b=\kappa_\tau$ and $\kappa_V^{} = \kappa_W^{}=\kappa_Z^{}$ are assumed. 
It is seen that both measured values of $\kappa_V^{}$ and 
$\kappa_f^{}$ agree with the corresponding SM predictions, i.e., 
$\kappa_V^{}=\kappa_f=1$ are included within the 2$\sigma$ level. 
However, the current 1$\sigma$ uncertainties, typically of ${\cal O}(10\%)$,  are not so small.
In future collider experiments, these scaling factors are expected to be measured more precisely. 
For example, $\kappa_V^{}$ will be measured with a few percent and less than 1 percent accuracy at the High-Luminosity LHC with $\sqrt{s}=14$ TeV and 
$3000$ fb$^{-1}$ of the integrated luminosity and 
at the International Linear Collider (ILC) with $\sqrt{s}=500$ GeV and $500$ fb$^{-1}$ of the integrated luminosity, respectively~\cite{snowmass}.

\begin{table}[ht]
\begin{center}
\caption{Simple extended Higgs sectors with $\rho_{\text{tree}}=1$ and without tree level FCNCs, where $\Phi$ denotes 
the isospin doublet with $Y=1/2$, and $\varphi(T,Y)$ denotes an extra scalar field with the isospin $T$ and the hypercharge $Y$. 
THDMs are further classified into four models with Type-I, Type-II, Type-X and Type-Y Yukawa interactions defined in Table~\ref{types}. }
\begin{tabular}{|c|l|l|l|l|}\hline
Models & Doublet-Singlet Model & THDMs & GM Model & Doublet-Septet Model   \\ \hline 
Scalar field contents & $\Phi+\varphi(0,0)$ & $\Phi+\varphi(1/2,1/2)$ & $\Phi+\varphi(1,0)+\varphi(1,1)$ & $\Phi+\varphi(3,2)$   \\\hline
\end{tabular}
\label{s}
\end{center}
\end{table}

In the following, we discuss the Higgs boson couplings in seven extended Higgs models with $\rho_{\text{tree}}=1$ and without FCNCs at the tree level. 
The scalar field contents in these models are shown in Table~\ref{s}.
In general, the THDMs have a CP-violating coupling in their scalar potential, but we neglect it for simplicity. 
Only in the GM model, we need a special treatment, because 
there are three CP-even scalar components, and extra triplet fields $\varphi(1,1)$ and $\varphi(1,0)$  
are introduced as a $3$ dimensional representation of the global $SU(2)_L\times SU(2)_R$ symmetry~\cite{GM}. 

First of all, we define the VEV $v$ related to the Fermi constant $G_F$ by $v^2=(\sqrt{2}G_F)^{-1}$ as follows
\begin{align}
v^2  = v_0^2 + \eta_{\text{ext}}^2 v_{\text{ext}}^2, \label{eta}
\end{align}
where $v_0$ and $v_{\text{ext}}$ are respectively the VEVs of $\Phi$ and the extra scalar field, i.e., 
$v_0\equiv\sqrt{2}\langle \Phi^0\rangle~~\text{and}~~v_{\text{ext}}\equiv \sqrt{2}\langle \varphi^0(T,Y)\rangle$. 
In the GM model, $v_{\text{ext}}$ is given as 
$\langle \varphi^0(1,1)\rangle=\langle \varphi^0(1,0)\rangle =  v_{\text{ext}}$. By taking this alignment, 
the $SU(2)_L\times SU(2)_R$ symmetry is spontaneously broken down into the custodial $SU(2)_V$ symmetry, and then $\rho_{\text{tree}}=1$ is satisfied. 
In Eq.~(\ref{eta}), the values of $\eta_{\text{ext}}$ are shown in Table~\ref{Tab:ScalingFactor}. 
In terms of $v_0$, $v_{\text{ext}}$ and $\eta_{\text{ext}}$, we can define generalized $\tan\beta$ which is usually introduced as the ratio of two VEVs in THDMs as follows
\begin{align}
\tan\beta =  v_0/(\eta_\text{ext}\, v_\text{ext}). 
\end{align}
Because the singlet field does not contribute to the electroweak symmetry breaking, we have $\eta_{\text{ext}}=0$, and we cannot define $\tan\beta$ in the Doublet-Singlet model. 

Next, we define the SM-like Higgs boson $h$. 
Let $h_0$ and $h_{\text{ext}}$ be the neutral CP-even components 
of $\Phi$ and that of the extra scalar field in the models excepted for the GM model, respectively. 
We then define $h$ by introducing a mixing angle $\alpha$ as
\begin{align}
\begin{pmatrix}
 h_\text{ext}^{}\\
 h_0
 \end{pmatrix}
=
\begin{pmatrix}
\cos\alpha & -\sin\alpha \\
\sin\alpha & \cos\alpha
\end{pmatrix}
\begin{pmatrix} 
H \\ 
h \end{pmatrix}, 
\end{align}
where $H$ is an extra CP-even Higgs boson. 
In the GM model, there are three CP-even scalar states from  $\varphi(1,1)$ and $\varphi(1,0)$ in addition to $h_0$. 
%
The mass eigenstates are given by the following transformation~\cite{yagyu}:
\begin{align}
\left(
\begin{array}{c}
h_0\\
\varphi^0(1,0)\\
\sqrt{2}\text{Re}\varphi^0(1,1)
\end{array}\right)=
\left(
\begin{array}{ccc}
1 & 0 &0\\
0 & \frac{1}{\sqrt{3}} & -\sqrt{\frac{2}{3}}\\
0 & \sqrt{\frac{2}{3}} & \frac{1}{\sqrt{3}}
\end{array}\right)
\left(
\begin{array}{ccc}
 \cos\alpha & \sin\alpha &0\\
 -\sin\alpha & \cos\alpha &0\\
0 & 0 & 1 
\end{array}\right)
\left(
\begin{array}{c}
h\\
H_1^0\\
H_5^0
\end{array}\right), 
\end{align}
where $H_5^0$ is the neutral $SU(2)_V$ 5-plet Higgs boson, and $H_1^0$ and $h$ are the $SU(2)_V$ singlets in the classification of $3\otimes 3=5\oplus 3\oplus 1$ 
under $SU(2)_V$. 

\begin{table}[ht]
\begin{center}
\caption{$\eta_{\text{ext}}$, $\tan\beta$ and the scaling factors $\kappa_f$ and $\kappa_V$ in the extended Higgs sectors given in Table~\ref{s}. 
The $\xi_f$ factors in $\kappa_f$ for THDMs are given in Table~\ref{types}.  }
\begin{tabular}{|c|l|l|l|l|}\hline
&$\eta_{\text{ext}}$ & $\tan\beta$ & $\kappa_f$ &$\kappa_V^{}$   \\ \hline 
Doublet-Singlet Model&0 & --- & $\cos\alpha$ & $\cos\alpha$ \\ \hline 
THDMs&1  &$v_0/v_\text{ext}^{}$ &$\sin(\beta-\alpha)+\xi_f\cos(\beta-\alpha)$ &$\sin(\beta-\alpha)$  \\ \hline 
GM Model&$2\sqrt{2}$ &$v_0/(2\sqrt{2}v_\text{ext}^{})$& $\cos\alpha/\sin\beta$ &   $\sin\beta \cos\alpha -\frac{2\sqrt{2}}{\sqrt{3}} \cos\beta \sin\alpha$ \\  \hline 
Doublet-Septet Model&$4$ &$v_0/(4v_\text{ext}^{})$&$\cos\alpha/\sin\beta$&$\sin\beta \cos\alpha 
-4 \cos\beta \sin\alpha$\\\hline
\end{tabular}
\end{center}
\label{Tab:ScalingFactor}
\end{table}

Now, we can express $\kappa_V^{}$ and $\kappa_f$ in all the models in terms of $\beta$ and $\alpha$ as 
shown in Table~\ref{Tab:ScalingFactor}. 
Let us summarize remarkable points for $\kappa_V^{}$ and $\kappa_f$ as follows:
\begin{enumerate}
\item We can classify all the models into two classes, (i) models with universal $\kappa_f$ ($\kappa_u=\kappa_d=\kappa_e$), i.e.,  
the Doublet-Singlet model,  the Type-I THDM, the GM model and the Doublet-Septet model, and 
(ii) those with non-universal $\kappa_f$, i.e., the Type-II, Type-X and Type-Y THDMs. 
\item
In the Doublet-Singlet model, both $\kappa_V$ and $\kappa_f$ are universally suppressed by $\cos\alpha$.
\item
In the THDMs, by taking the limit $\sin(\beta-\alpha)\to 1$~\cite{Haber}, both $\kappa_V$ and $\kappa_f$ become unity. 
In the other models, the similar limit is defined by $\alpha=\beta=0$.  
\item
In the GM model and the Doublet-Septet model, $\kappa_V$ can be greater than 1~\cite{kv1,kv2}. 
This feature is originated by  $\eta_{\text{ext}}>1$.  
\end{enumerate}

\begin{figure}[ht]
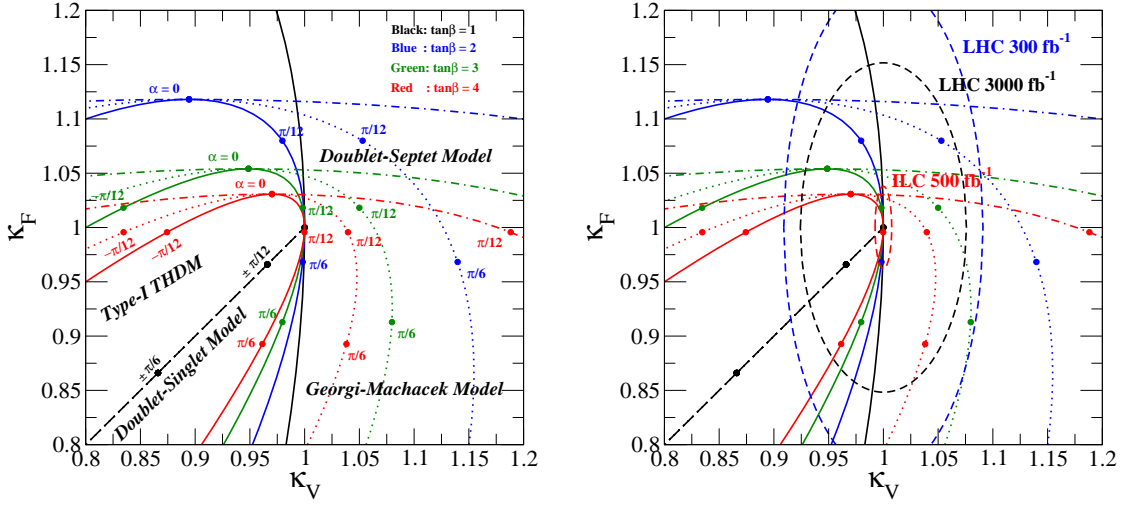

\centering
\includegraphics[width=70mm]{KV-KF_ALL2.eps}\hspace{5mm}
\includegraphics[width=70mm]{KV-KF_ALL3.eps}
\caption{
The scaling factors $\kappa_f$ and $\kappa_V$ for each value of $\beta$ and $\alpha$ in the Doublet-Singlet model (dashed), 
the Type-I THDM (solid), the GM model (dotted), and the Doublet-Septet model (dash-dotted)~\cite{fingerprint}.
The right panel also shows the expected 1$\sigma$ uncertainty for the measurements of $\kappa_f$ and $\kappa_V$ at the 
LHC with 300 fb$^{-1}$ and $3000$ fb$^{-1}$, and at the ILC with $\sqrt{s}=$500 GeV and 500 fb$^{-1}$~\cite{snowmass}. } 
\label{fig1}
\end{figure}

\begin{figure}[ht]
\centering
\includegraphics[width=70mm]{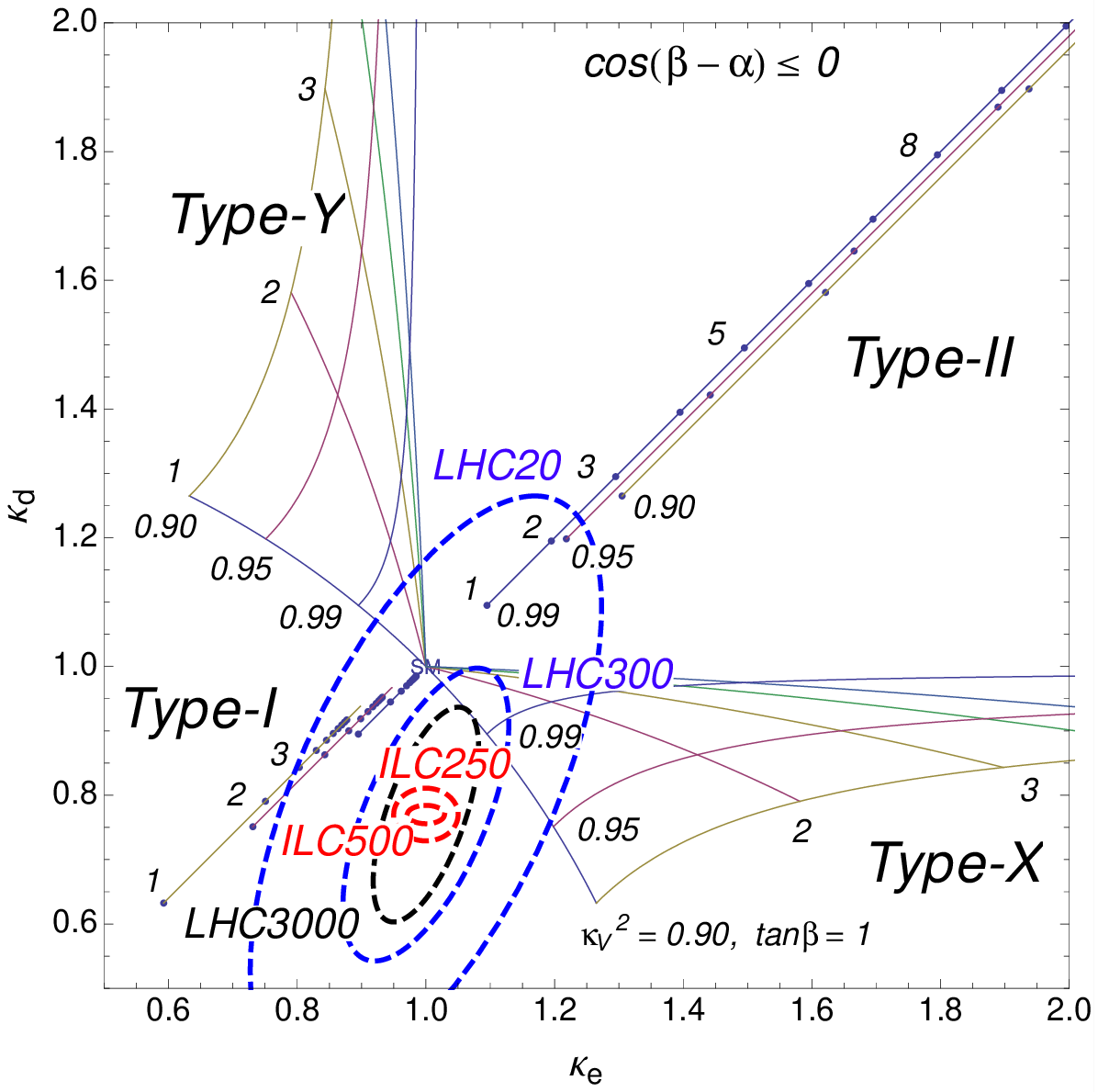}
\includegraphics[width=70mm]{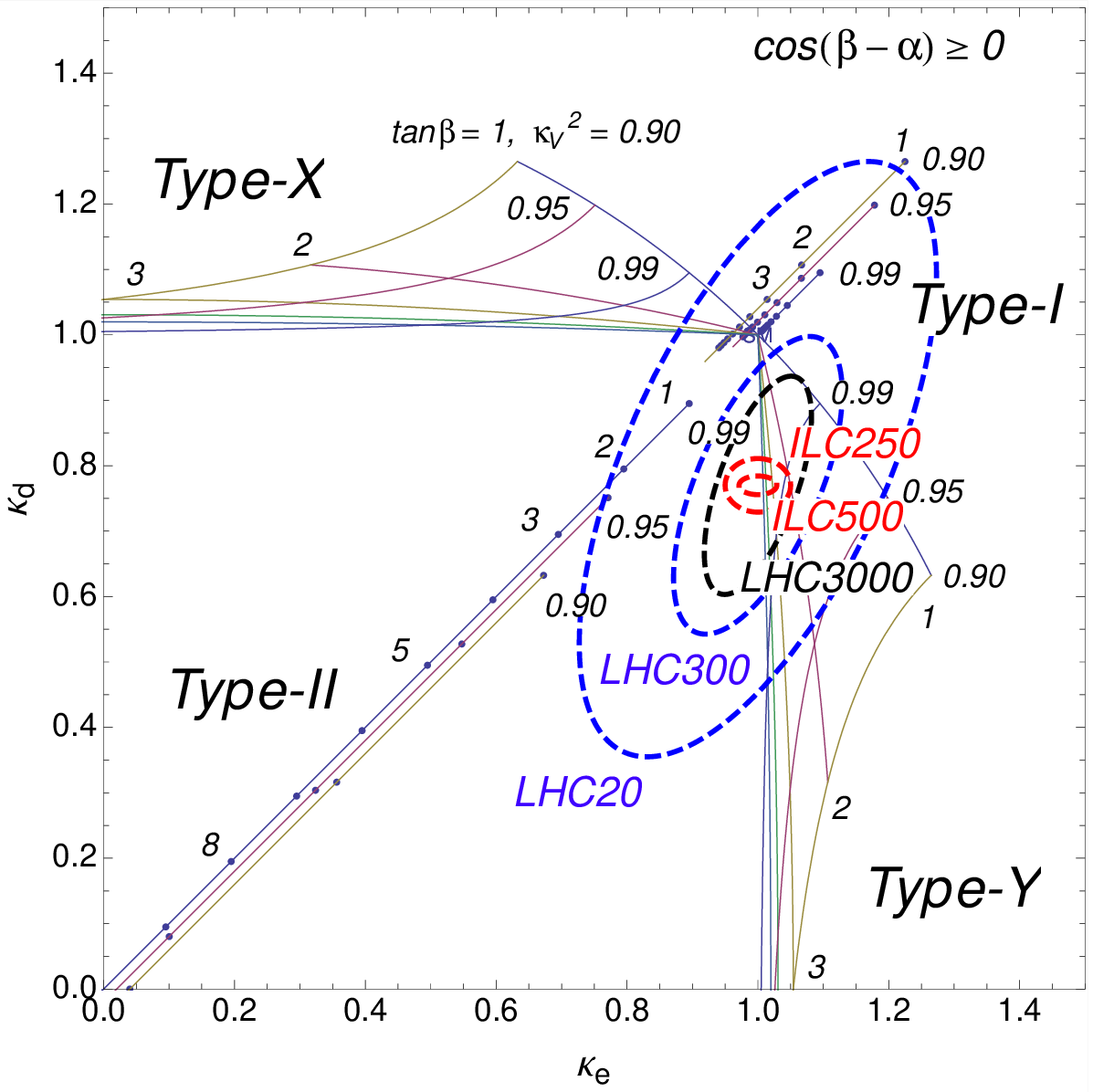}
\caption{The flavor universal scaling factors $\kappa_e$ and $\kappa_d$ for each value of $\beta$ and $\alpha$ in the four THDMs with different types of Yukawa interactions~\cite{fingerprint}.
The left (right) panel shows the case of $\cos(\beta-\alpha)<0$ ($\cos(\beta-\alpha)>0$). 
} \label{fig2}
\end{figure}

In FIG.~\ref{fig1}, we show predictions of $\kappa_V$ and $\kappa_f (=\kappa_u=\kappa_d=\kappa_e)$ for each value of 
$\alpha$ and $\beta$ in the models classified as (i) in the above. 
In the right panel, the expected 1$\sigma$ uncertainty for the measurements of $\kappa_f$ and $\kappa_V$ at the LHC and the ILC are also shown in addition to those 
predictions. 
As mentioned in 4. in the above, 
the GM model and the Doublet-Septet model provide $\kappa_V^{}>1$ in some parameter choices, 
which is not allowed in the other models. 
Therefore, $\kappa_V>1$ can be a smoking gun to prove these models.  

In FIG.~\ref{fig2}, we show predictions of flavor independent scaling factors on the $\kappa_e$ and $\kappa_d$ plane in the four THDMs  with different types of Yukawa interactions. 
It is clearly seen that the THDMs with different types give a prediction in the different quadrants.  
Therefore, precise measurements of $\kappa_e$ and $\kappa_d$ can be a powerful tool to determine the type of Yukawa interactions. 

\section{Conclusions and Discussions}

We have discussed deviations in the SM-like Higgs boson $h$ couplings with the gauge bosons ($hVV$) and fermions ($hf\bar{f}$) in several 
simple non-minimal Higgs sectors with $\rho_{\text{tree}}=1$ and without tree level FCNCs. 
We have shown that characteristic patterns of deviations in $hVV$ and $hf\bar{f}$ couplings appear in each of Higgs sectors as seen in FIGs.~\ref{fig1} and \ref{fig2}. 
Therefore, we can distinguish non-minimal Higgs sectors 
by {\it fingerprinting} the Higgs boson couplings, namely, 
comparing the predicted Higgs boson couplings and precisely measured values at future collider experiments. 

Although we have investigated the Higgs boson couplings at the tree level, studying radiative corrections to these are quite important, because of the two reasons: 
(i) we have to confirm how the tree level results as seen in FIGs.~\ref{fig1} and \ref{fig2} 
can be modified by loop corrections, and (ii) we can extract inner parameters in a model. 
Regarding (ii), when we compute the radiative corrections to the Higgs boson couplings, those magnitudes depend on not only tree level mixing parameters $\beta$ and $\alpha$
but also other parameters such as masses of extra Higgs bosons. 
Finally, we would like to introduce some works for radiative corrections to the $h$ couplings in non-minimal Higgs sectors. 
In non-supersymmetric THDMs, one-loop corrections to the $hVV$ and $hhh$ couplings have been calculated in Refs.~\cite{KKOSY,KOSY}, 
and those to $hf\bar{f}$ have been computed in \cite{KKY}. 
Recently, extractions of inner parameters in the THDMs have been discussed in Ref.~\cite{Yagyu_THDM}.  
For the Higgs sector in the minimal supersymmetric SM, radiative corrections to the Higgs boson couplings have been calculated in~\cite{thdm_rad_susy}.  
The one-loop calculations for the Higgs couplings in the Higgs triplet model have also been calculated in Refs.~\cite{AKKY_Lett,AKKY_Full}.


\begin{acknowledgments}
I would like to thank Shinya Kanemura, Koji Tsumura and Hiroshi Yokoya for fruitful collaborations. 
I am also grateful to all the organizers of the conference HPNP2015 at University of Toyama, especially for Shinya Kanemura, for their warm
hospitality and giving me the opportunity to present a talk.
This work was supported by JSPS postdoctoral fellowships for research abroad.
\end{acknowledgments}

\bigskip 

\end{document}